\documentclass[showpacs,aps,prd]{revtex4}
\input epsf

\begin{document}

\title{$S$-wave $D^{(*)}N$ molecular states: $\Sigma_{c}(2800)$ and $\Lambda_{c}(2940)^{+}$?}
\author{Jian-Rong Zhang}
\affiliation{Department of Physics, College of Science, National University of Defense Technology,
Changsha 410073, Hunan, People's Republic of China}

\begin{abstract}
Theoretically, some works have proposed the hadronic
resonances $\Sigma_{c}(2800)$ and $\Lambda_{c}(2940)^{+}$ to be $S$-wave
$DN$ and $D^{*}N$ molecular candidates, respectively.
In the framework of QCD sum rules, we investigate that whether $\Sigma_{c}(2800)$ and $\Lambda_{c}(2940)^{+}$
could be explained as the $S$-wave $DN$ state with $J^{P}=\frac{1}{2}^{-}$ and the $S$-wave $D^{*}N$ state with $J^{P}=\frac{3}{2}^{-}$,
respectively.
Technically, contributions of operators up to dimension $12$ are included in the
operator product expansion (OPE).
The final results are $3.64\pm0.33~\mbox{GeV}$ and $3.73\pm0.35~\mbox{GeV}$
for the $S$-wave $DN$ state of $J^{P}=\frac{1}{2}^{-}$
 and the $S$-wave $D^{*}N$ state of $J^{P}=\frac{3}{2}^{-}$, respectively.
They are somewhat bigger than the experimental data of $\Sigma_{c}(2800)$ and $\Lambda_{c}(2940)^{+}$,
respectively.
In view of that corresponding molecular currents are constructed from local operators
of hadrons, the possibility of $\Sigma_c(2800)$ and $\Lambda_{c}(2940)^{+}$ as molecular states
can not be arbitrarily excluded merely from these disagreements
between molecular masses using local currents and experimental data.
But then these results imply that $\Sigma_{c}(2800)$ and $\Lambda_{c}(2940)^{+}$ could not be compact states. 
This may suggest a limitation of the QCD sum rule using the local current to determine whether some state is a molecular state or not.
As byproducts, masses for their
bottom partners are predicted to be $6.97\pm0.34~\mbox{GeV}$ for the $S$-wave $\bar{B}N$ state of $J^{P}=\frac{1}{2}^{-}$ and
$6.98\pm0.34~\mbox{GeV}$ for the $S$-wave $\bar{B}^{*}N$ state of $J^{P}=\frac{3}{2}^{-}$.
\end{abstract}
\pacs {11.55.Hx, 12.38.Lg, 12.39.Mk}\maketitle

\section{Introduction}\label{sec1}
In the past several years, many new excited charmed baryonic states have been discovered experimentally.
For example, Belle Collaboration observed
an isotriplet of new states $\Sigma_{c}(2800)$
decaying into $\Lambda_{c}^{+}\pi$, and they
tentatively
identified the quantum numbers of these states as $J^{P}=\frac{3}{2}^{-}$ \cite{2800-belle}.
In particular,
the neutral state
$\Sigma_{c}(2800)^{0}$ was possibly confirmed in the $B^{-}\rightarrow\Sigma_{c}(2800)^{0}\bar{p}$ channel by Babar Collaboration  \cite{2800-babar}.
However, the measured mass $2846\pm8\pm10~\mbox{MeV}$ for $\Sigma_{c}(2800)^{0}$
 is $3\sigma$ higher (assuming Gaussian statistics) than the Belle's
measured value,
and Babar indicated that there is weak evidence that the excited $\Sigma_{c}^{0}$ observed by them is
$J=\frac{1}{2}$.
Moreover, Babar collaboration reported
the observation of a new
charmed state $\Lambda_{c}(2940)^{+}$ decaying to $D^{0}p$ with a mass of
$2939.8\pm1.3\pm1.0
~\mbox{MeV}$ and an intrinsic width of $17.5\pm5.2\pm5.9~\mbox{MeV}$ \cite{2940-babar}.
Subsequently, Belle Collaboration
confirmed it in the $\Lambda_{c}(2940)^{+}\rightarrow\Sigma_{c}(2455)^{0,++}\pi^{+,-}$
decay and measured its mass and width to be $2938.0\pm1.3^{+2.0}
_{-4.0}~\mbox{MeV}$ and $13^{+8+27}_{-5-7}
~\mbox{MeV}$, respectively \cite{2940-belle}.

The experimental observations have triggered theorists' great
interest in understanding the internal structures of $\Sigma_{c}(2800)$ and $\Lambda_{c}(2940)^{+}$.
One direct way of theoretical studies grounds on the
assignments of them as conventional charmed baryons.
From a relativized potential model prediction,
masses of $\Sigma_{c}(2800)$ and $\Lambda_{c}(2940)^{+}$
are close to theoretical values of $\Sigma_{c}^{*}$ with $J^{P}=\frac{3}{2}^{-}$ or $\frac{5}{2}^{-}$ and $\Lambda_{c}^{*}$ with $J^{P}=\frac{5}{2}^{-}$ or $\frac{3}{2}^{+}$, respectively \cite{quark-model}.
In the relativistic quark-diquark
picture, Ebert {\it et al.} suggested $\Sigma_{c}(2800)$
as one of the orbital ($1P$) excitations
of the $\Sigma_{c}$ with $J^{P}=\frac{1}{2}^{-}$, $\frac{3}{2}^{-}$, or $\frac{5}{2}^{-}$,
and proposed $\Lambda_{c}(2940)^{+}$ as
the first radial excitation of $\Sigma_{c}$ with $J^{P}=\frac{3}{2}^{+}$ \cite{Ebert}.
Strong decays of $\Sigma_{c}(2800)$ and $\Lambda_{c}(2940)^{+}$ as charmed baryons were analyzed
in heavy baryon chiral perturbation theory \cite{Cheng},
with the $^{3}P_{0}$ model \cite{Chen}, and using the chiral quark model \cite{Zhong}.
In Ref. \cite{Garcilazo}, Garcilazo {\it et al.} indicated that
$\Sigma_{c}(2800)$
would correspond to an orbital excitation
with $J^{P}=\frac{1}{2}^{-}$ or $\frac{3}{2}^{-}$
and
$\Lambda_{c}(2940)^{+}$ may
constitute the second orbital excitation of the
$\Lambda_{c}$ by the Faddeev
method. In a mass loaded flux tube
model, Chen {\it et al.} suggested that $\Lambda_{c}(2940)^{+}$
could be the orbitally excited $\Lambda_{c}^{+}$ with $J^{P}=\frac{5
}{2}^{-}$ \cite{BChen}. He {\it et al.}
evaluated the production rate of $\Lambda_{c}(2940)^{+}$ as a charmed baryon
at PANDA \cite{JHe}.

Another different way bases on the assumption that $\Sigma_{c}(2800)$ and $\Lambda_{c}(2940)^{+}$ are some molecular candidates.
Lutz {\it et al.} interpreted $\Sigma_{c}(2800)$ as a chiral molecule \cite{Lutz}.
In Ref. \cite{Tejero}, $\Sigma_{c}(2800)$ was deciphered as a dynamically generated resonance with a dominant $DN$
configuration. Dong {\it et al.} pursued a possible hadronic molecule interpretation of $\Sigma_{c}(2800)$ as a
bound state of the
charmed $D$ meson and the nucleon $N$, since isotriplet states of $\Sigma_{c}(2800)$ are very close to
respective $DN$ thresholds \cite{Dong-2800}.
They chose several possible
quantum number assignments of $\Sigma_{c}(2800)$ as $J^{P}=\frac{1}{2}^{\pm}$ and $\frac{3}{2}^{\pm}$.
Here $J^{P}=\frac{1}{2}^{-}$ corresponds to a $S$-wave $DN$ configuration, $J^{P}=\frac{1}{2}^{+}$ and $J^{P}=\frac{3}{2}^{+}$ represent
a $P$-wave, and $J^{P}=\frac{3}{2}^{-}$ has a relative $D$-wave in the $DN$ system.
They finally concluded that $\Sigma_{c}\rightarrow\Lambda_{c}\pi$
decay widths are consistent with current data for $J^{P}=\frac{1}{2}^{+}$
and $J^{P}=\frac{3}{2}^{-}$
assignments.
Coming down to $\Lambda_{c}(2940)^{+}$, it was
firstly proposed to be a $S$-wave $D^{*0}p$ molecular state
with $J^{P}=\frac{1}{2}^{-}$ in Ref. \cite{He},
because its mass is just a few $\mbox{MeV}$ below the $D^{*0}p$ threshold.
From an effective Lagrangian approach,
the strong two-body decay of $\Lambda_{c}(2940)^{+}$ was studied
under the $J^{P}=\frac{1}{2}^{-}$
and $\frac{1}{2}^{+}$  $D^{*}N$ molecular assignments, and it showed that
$J^{P}=\frac{1}{2}^{-}$ should be ruled out \cite{Dong}.
Later, the radiative and strong three-body
decays of $\Lambda_{c}(2940)^{+}$ were also researched in the $D^{*}N$ molecule picture
with $J^{P}=\frac{1}{2}^{+}$
\cite{Dong1,Dong2}.
He {\it et al.} systematically
studied the interaction between $D^{*}$ and $N$, and concluded
that the $D^{*}N$ systems may behave as $J^{P}=\frac{1}{2}^{\pm}$ and $\frac{3}{2}^{\pm}$
states \cite{JHe2}. In Ref. \cite{Recio}, Garc\'{\i}a-Recio {\it et al.} found a possible molecular candidate for the
$\Lambda_{c}(2940)^{+}$ in the $\frac{3}{2}^{-}$ channel.
Ortega {\it et al.} studied $\Lambda_{c}(2940)^{+}$ as a $D^{*}N$ molecule
with $J^{P}=\frac{3}{2}^{-}$ in a constituent quark model, and claimed
obtaining a mass which agrees with the experimental data \cite{Ortega}.

Although various interpretations to $\Sigma_{c}(2800)$ and $\Lambda_{c}(2940)^{+}$
were put forward, at present their underlying structures are still unclear, which means that it is
interesting and significative to make more theoretical
efforts to reveal their properties. Therefore, we devote to studying that whether $\Sigma_{c}(2800)$ and $\Lambda_{c}(2940)^{+}$
could be the $S$-wave $DN$ state with $J^{P}=\frac{1}{2}^{-}$ and the $S$-wave $D^{*}N$ state with $J^{P}=\frac{3}{2}^{-}$,
respectively.
QCD is widely believed nowadays to be a true theory of strong interactions.
At high energy, the effective coupling constant of the quark-gluon
interaction becomes small because of asymptotic freedom and the interaction can be treated perturbatively.
On the other hand, quark interaction within hadrons is strong since
it binds quarks into unseparable pairs. Thus,
low energy QCD involves a regime where it is futile to attempt perturbative calculations,
and the strong interaction dynamics of hadronic systems
is governed by nonperturbative QCD effects completely.
There are still many questions on nonperturbative QCD remain
unanswered or realized only at a qualitative level
since one's absence of knowledge on QCD confinement effects.
Therefore, it is quite difficult to
calculate the hadron spectrum from QCD first principles.
The method of
QCD sum rules \cite{svzsum}, developed by Shifman, Vainshtein, and Zakharov,
represents an attempt to bridge the gap between
the perturbative and nonperturbative sectors by employing the language
of dispersion relations.
It is well known for
the advantages of this method:
instead of a model-dependent treatment in terms of constituent quarks,
hadrons are represented by their interpolating quark currents and
the interactions of quark-gluon currents with QCD
vacuum fields critically depend on the quantum numbers (spin-parity, flavor content)
of these currents.
The QCD sum rule method is a nonperturbative
formulation firmly based on the first principle of QCD,
which
has become a widely used working tool in hadron phenomenology.
The mere fact that the seminal paper on QCD sum rules have already been
cited more than $4000$ times reflects its
vigorousness.
It has been successfully applied
to conventional mesons and baryons (for reviews
see \cite{overview1,overview2,overview3,overview4} and references
therein) and multiquark states (e.g.
see \cite{XYZ}).
In particular,
many theoretical practitioners began to study
light pentaquark states in Refs. \cite{pentaquark,pentaquark1} and
heavy pentaquark systems in Ref. \cite{pentaquark-heavy}.

In this work, we intend to
investigate that whether $\Sigma_{c}(2800)$ and $\Lambda_{c}(2940)^{+}$
could be the $S$-wave $DN$ state with $J^{P}=\frac{1}{2}^{-}$
and the $S$-wave $D^{*}N$ state with $J^{P}=\frac{3}{2}^{-}$
respectively from QCD sum rules.
The rest of the paper is organized as follow. We derive QCD
sum rules for molecular states in Sec. \ref{sec2}, with similar techniques
as our previous works on heavy baryons \cite{Zhang} and molecular states \cite{Zhang1}.
The numerical analysis and discussions are presented in Sec. \ref{sec3},
and masses of $D^{(*)}N$ and $\bar{B}^{(*)}N$ molecular states are extracted
out. Sec. \ref{sec4} contributes to the conclusions.

\section{QCD sum rules for meson-nucleon molecular states}\label{sec2}
\subsection{constructions of interpolating currents}
One basic point of QCD sum rules is to construct a proper
interpolating current to represent the studied state.
In the real world, one hadron in particular a molecular state can not be an ideal point
particle in a rigorous manner
because each constituent
quark of a hadronic system is separated in the space.
Without doubt, it would be best if one could describe a real hadron
using some nonlocal current in QCD sum rules. However, the practitioners can find that
it would become quite difficult or even unfeasible for QCD sum rule calculations when a
hadron's current is constructed nonlocal. Thus,
interpolating currents used
in QCD sum rules are commonly built local to
characterize real hadrons, which is in fact a limitation inherent in the QCD sum rule method
disposal of hadrons.
The simplification has been widely proved feasible
and the QCD sum rule method has been successfully applied to plenty of hadrons, involving
a number of works on molecular states  since the observations of so-called ``X", ``Y", and ``Z"
new hadrons in recent years (e.g. see \cite{XYZ} and references
therein).
Following the usual treatment, in this work we will construct molecular currents from local operators
of hadrons.
At present, molecular currents are built up with the color-singlet currents of
composed hadrons to form hadron-hadron
configurations of fields, which are different
from currents of pentaquark states constructed by diquark-diquark-antiquark configurations
of fields. Although molecular currents can be related to
pentaquark currents by Fierz rearrangement, the transformation relations are suppressed by corresponding color and Dirac
factor. Consequently, it would be best to choose a hadron-hadron type of current to characterize if the studied object is a molecular state.

Therefore, currents for $S$-wave
 $D^{(*)}N$ or $\bar{B}^{(*)}N$ molecular states
can be  built up with the color-singlet currents of
$D^{(*)}$ or $\bar{B}^{(*)}$ mesons and $N$ nucleons to form meson-nucleon configurations of fields.
In full theory, interpolating currents for $D^{(*)}$ and $\bar{B}^{(*)}$
mesons can be found in Ref. \cite{reinders}, and
currents for nucleons have been listed in Ref. \cite{baryon-current}.
Therefore, we build following forms of currents:
\begin{eqnarray}
j&=&(\bar{q}^{c^{'}}i\gamma_{5}Q^{c^{'}})(\varepsilon_{abc}q_{1}^{Ta}C\gamma_{\mu}q_{2}^{b}\gamma_{5}\gamma^{\mu}q_{3}^{c}),
\end{eqnarray}
for the $S$-wave $DN$ or $\bar{B}N$ molecular state with $J^{P}=\frac{1}{2}^{-}$, and
\begin{eqnarray}
j^{\rho}&=&(\bar{q}^{c^{'}}\gamma^{\rho}Q^{c^{'}})(\varepsilon_{abc}q_{1}^{Ta}C\gamma_{\mu}q_{2}^{b}\gamma_{5}\gamma^{\mu}q_{3}^{c}),
\end{eqnarray}
for the $S$-wave
 $D^{*}N$ or $\bar{B}^{*}N$ molecular state with $J^{P}=\frac{3}{2}^{-}$.
Here $Q$ is heavy quark $c$ or $b$, and $q_{1}$, $q_{2}$, as well as $q_{3}$ denote light quarks $u$ and/or $d$.
The index $T$ means matrix
transposition, $C$ is the charge conjugation matrix,
with $a$, $b$, $c$ and $c'$ as color indices.
Beside $J^{P}=\frac{3}{2}^{-}$, one may have noted that
the quantum number for a $S$-wave $D^{*}N$ molecule could also be $\frac{1}{2}^{-}$.
However, it is not straightforward to construct the interpolating current
for the $S$-wave $D^{*}N$ molecule with $J^{P}=\frac{1}{2}^{-}$
from meson-nucleon configurations of fields.
That's the main reason why the $S$-wave $D^{*}N$ molecule with $J^{P}=\frac{1}{2}^{-}$
has not been involved here.
In addition, it showed that the molecular assignment of $\Lambda_{c}(2940)^{+}$ with
$J^{P}=\frac{1}{2}^{-}$ should be ruled out
from an effective Lagrangian approach \cite{Dong}.

\subsection{QCD sum rules for meson-nucleon molecular states}
QCD sum rules for $DN$ and $\bar{B}N$ molecular states with $J^{P}=\frac{1}{2}^{-}$ are constructed from the two-point correlator
\begin{eqnarray}\label{correlator}
\Pi(q^{2})=i\int
d^{4}x\mbox{e}^{iq.x}\langle0|T[j(x)\overline{j}(0)]|0\rangle.
\end{eqnarray}
Lorentz covariance hints that the
correlator has the
form
\begin{eqnarray}
\Pi(q^{2})=\rlap/q\Pi_{1}(q^{2})+\Pi_{2}(q^{2}).
\end{eqnarray}
Phenomenologically, the correlator can be
expressed as
\begin{eqnarray}\label{pole-model}
\Pi(q^{2})=\lambda^{2}_H\frac{\rlap/q-M_{H}}{M_{H}^{2}-q^{2}}+\frac{1}{\pi}\int_{s_{0}}
^{\infty}ds\frac{\rlap/q\mbox{Im}\Pi_{1}^{\mbox{phen}}(s)+\mbox{Im}\Pi_{2}^{\mbox{phen}}(s)}{s-q^{2}}+...,
\end{eqnarray}
where $M_{H}$ is the mass of the hadronic resonance, $s_0$ is the threshold parameter, and
$\lambda_{H}$ gives the coupling of the current to the hadron
$\langle0|j|H\rangle=\lambda_{H}v(p,s)$. In the OPE side,
the correlator can be written as
\begin{eqnarray}
\Pi(q^{2})=\rlap/q\bigg\{\int_{m_{Q}^{2}}^{\infty}ds\frac{\rho_{1}(s)}{s-q^{2}}+\Pi_{1}^{\mbox{cond}}(q^{2})\bigg\}+\int_{m_{Q}^{2}}^{\infty}ds\frac{\rho_{2}(s)}{s-q^{2}}+\Pi_{2}^{\mbox{cond}}(q^{2}),
\end{eqnarray}
After equating the two sides for $\Pi(q^{2})$, assuming
quark-hadron duality, making a Borel transform, and
transferring the continuum contribution to the OPE side, the sum
rules can be written as
\begin{eqnarray}\label{sr1}
\lambda^{2}_He^{-M_{H}^{2}/M^{2}}&=&\int_{m_{Q}^{2}}^{s_{0}}ds\rho_{1}(s)e^{-s/M^{2}}+\hat{B}\Pi_{1}^{\mbox{cond}},
\end{eqnarray}
\begin{eqnarray}\label{sr2}
-\lambda^{2}_HM_{H}e^{-M_{H}^{2}/M^{2}}&=&\int_{m_{Q}^{2}}^{s_{0}}ds\rho_{2}(s)e^{-s/M^{2}}+\hat{B}\Pi_{2}^{\mbox{cond}},
\end{eqnarray}
where $M^2$ indicates the Borel parameter.

There is some difference while deriving
mass sum rules for $D^{*}N$ and $\bar{B}^{*}N$ molecular states with $J^{P}=\frac{3}{2}^{-}$.
One can start from the two-point correlator
\begin{eqnarray}\label{correlator2}
\Pi^{\rho\tau}(q^{2})=i\int
d^{4}x\mbox{e}^{iq.x}\langle0|T[j^{\rho}(x)\overline{j^{\tau}}(0)]|0\rangle.
\end{eqnarray}
Lorentz covariance implies that the
two-point correlator in Eq. (\ref{correlator2}) has the
form
\begin{eqnarray}
\Pi^{\rho\tau}(q^{2})=-g^{\rho\tau}[\rlap/q\Pi_{1}(q^{2})+\Pi_{2}(q^{2})]+...,
\end{eqnarray}
where the ellipse denotes other Lorentz structures which acquire contributions
from both $J=\frac{1}{2}$ and $J=\frac{3}{2}$. The tensor
structures $g^{\rho\tau}\rlap/q$ and $g^{\rho\tau}$
are contributed only by the $J=\frac{3}{2}$ hadrons.
The phenomenological side of $\Pi^{\rho\tau}(q^{2})$ can be expressed as
\begin{eqnarray}\label{pole-model}
\Pi^{\rho\tau}(q^{2})=-g^{\rho\tau}\Bigg\{\lambda_{H}^{2}\frac{\rlap/q-M_{H}}{M_{H}^{2}-q^{2}}+\frac{1}{\pi}\int_{s_{0}}
^{\infty}ds\frac{\rlap/q\mbox{Im}\Pi_{1}^{\mbox{phen}}(s)+\mbox{Im}\Pi_{2}^{\mbox{phen}}(s)}{s-q^{2}}\Bigg\}+...,
\end{eqnarray}
where $\lambda_{H}$ gives the
coupling of the hadronic state to the current $j^{\rho}$ as
$\langle0|j^{\rho}|H\rangle=\lambda_{H}v^{\rho}(q,s)$.
Here, $v^{\rho}(q,s)$ is the Rarita-Schwinger spinor for
$J^{P}=\frac{3}{2}^{-}$.
In the OPE side, one can write
the correlator as
\begin{eqnarray}
\Pi^{\rho\tau}(q^{2})=-g^{\rho\tau}\Bigg\{\rlap/q\bigg[\int_{m_{Q}^{2}}^{\infty}ds\frac{\rho_{1}(s)}{s-q^{2}}+\Pi_{1}^{\mbox{cond}}(q^{2})\bigg]+\int_{m_{Q}^{2}}^{\infty}ds\frac{\rho_{2}(s)}{s-q^{2}}+\Pi_{2}^{\mbox{cond}}(q^{2})\Bigg\},
\end{eqnarray}
Equating the two sides for $\Pi^{\rho\tau}(q^{2})$, assuming
quark-hadron duality, and making a Borel transform, the sum
rules can be expressed as
\begin{eqnarray}\label{qsr1}
\lambda^{2}_He^{-M_{H}^{2}/M^{2}}&=&\int_{m_{Q}^{2}}^{s_{0}}ds\rho_{1}(s)e^{-s/M^{2}}+\hat{B}\Pi_{1}^{\mbox{cond}},
\end{eqnarray}
\begin{eqnarray}\label{qsr2}
-\lambda^{2}_H M_{H}e^{-M_{H}^{2}/M^{2}}&=&\int_{m_{Q}^{2}}^{s_{0}}ds\rho_{2}(s)e^{-s/M^{2}}+\hat{B}\Pi_{2}^{\mbox{cond}}.
\end{eqnarray}
To eliminate the hadron coupling constant $\lambda_H$ in sum rules
(\ref{sr1}), (\ref{sr2}), (\ref{qsr1}), and (\ref{qsr2}), one can take the derivatives of sum rules
with respect to $1/M^{2}$, divide the results by themselves to get
\begin{eqnarray}\label{sumrule1}
M_{H}^{2}&=&\bigg\{\int_{m_{Q}^{2}}^{s_{0}}ds\rho_{1}(s)s
e^{-s/M^{2}}+d/d(-\frac{1}{M^{2}})\hat{B}\Pi_{1}^{\mbox{cond}}\bigg\}/
\bigg\{\int_{m_{Q}^{2}}^{s_{0}}ds\rho_{1}(s)e^{-s/M^{2}}
+\hat{B}\Pi_{1}^{\mbox{cond}}\bigg\},
\end{eqnarray}
\begin{eqnarray}\label{sumrule2}
M_{H}^{2}&=&\bigg\{\int_{m_{Q}^{2}}^{s_{0}}ds\rho_{2}(s)s
e^{-s/M^{2}}+d/d(-\frac{1}{M^{2}})\hat{B}\Pi_{2}^{\mbox{cond}}\bigg\}/
\bigg\{\int_{m_{Q}^{2}}^{s_{0}}ds\rho_{2}(s)e^{-s/M^{2}}
+\hat{B}\Pi_{2}^{\mbox{cond}}\bigg\}.
\end{eqnarray}

\subsection{spectral densities}
The spectral density is given by the correlator's imaginary part
\begin{eqnarray}
\rho_{i}(s)=\frac{1}{\pi}\mbox{Im}\Pi_{i}^{\mbox{OPE}}(s),~~i=1,2.
\end{eqnarray}
In the concrete OPE calculation, one works at leading order in $\alpha_{s}$ and considers condensates up
to dimension $12$.
Note that $O(\alpha_s)$
corrections may be important in the QCD sum rule calculations.
Meanwhile, one could expect the calculations of $O(\alpha_s)$ corrections especially
for multiquark systems
are complicated and tedious as one has to deal
with many multi-loop massive propagator diagrams.
Actually, a lot of hard
calculations already need to be done even if one merely works at leading order since there
include many Feynman diagrams up to dimension $12$.
However, it is expected that the $O(\alpha_s)$ corrections might be
under control since a partial cancelation occurs in the ratio
obtaining the mass sum rules (\ref{sumrule1}) and (\ref{sumrule2}). This has been proved to
be true in the analysis for heavy mesons \cite{Narison}
(the value of $f_{D}$ increases by $12\%$ after the inclusion of the
$O(\alpha_s)$ correction) and singly heavy baryons \cite{Groote} (the
corrections increase the calculated baryon
masses by about $10\%$). Furthermore, in order to improve on the accuracy of the QCD sum rule
analysis for molecular states, one could take into
account the $O(\alpha_s)$
corrections in the further work after fulfilling a burdensome task.
To keep the heavy-quark mass finite, one can use the
momentum-space expression for the heavy-quark propagator \cite{reinders}
\begin{eqnarray}
S_{Q}(p)&=&\frac{i}{\rlap/p-m_{Q}}
-\frac{i}{4}gt^{A}G^{A}_{\kappa\lambda}(0)\frac{1}{(p^{2}-m_{Q}^{2})^{2}}[\sigma_{\kappa\lambda}(\rlap/p+m_{Q})
+(\rlap/p+m_{Q})\sigma_{\kappa\lambda}]\nonumber\\&&{}
-\frac{i}{4}g^{2}t^{A}t^{B}G^{A}_{\alpha\beta}(0)G^{B}_{\mu\nu}(0)\frac{\rlap/p+m_{Q}}{(p^{2}-m_{Q}^{2})^{5}}[
\gamma_{\alpha}(\rlap/p+m_{Q})\gamma_{\beta}(\rlap/p+m_{Q})\gamma_{\mu}(\rlap/p+m_{Q})\gamma_{\nu}\\&&{}
+\gamma_{\alpha}(\rlap/p+m_{Q})\gamma_{\mu}(\rlap/p+m_{Q})\gamma_{\beta}(\rlap/p+m_{Q})\gamma_{\nu}+
\gamma_{\alpha}(\rlap/p+m_{Q})\gamma_{\mu}(\rlap/p+m_{Q})\gamma_{\nu}(\rlap/p+m_{Q})\gamma_{\beta}](\rlap/p+m_{Q})\nonumber\\&&{}
+\frac{i}{48}g^{3}f^{ABC}G^{A}_{\gamma\delta}G^{B}_{\delta\varepsilon}G^{C}_{\varepsilon\gamma}\frac{1}{(p^{2}-m_{Q}^{2})^{6}}(\rlap/p+m_{Q})
[\rlap/p(p^{2}-3m_{Q}^{2})+2m_{Q}(2p^{2}-m_{Q}^{2})](\rlap/p+m_{Q}).\nonumber
\end{eqnarray}
The light-quark part of the
correlator can be calculated in the coordinate space employing the light-quark
propagator
\begin{eqnarray}
S_{ab}(x)&=&\frac{i\delta_{ab}}{2\pi^{2}x^{4}}\rlap/x-\frac{m_{q}\delta_{ab}}{4\pi^{2}x^{2}}-\frac{i}{32\pi^{2}x^{2}}t^{A}_{ab}gG^{A}_{\mu\nu}(\rlap/x\sigma^{\mu\nu}
+\sigma^{\mu\nu}\rlap/x)-\frac{\delta_{ab}}{12}\langle\bar{q}q\rangle+\frac{i\delta_{ab}}{48}m_{q}\langle\bar{q}q\rangle\rlap/x\nonumber\\&&{}\hspace{-0.3cm}
-\frac{x^{2}\delta_{ab}}{3\cdot2^{6}}\langle g\bar{q}\sigma\cdot Gq\rangle
+\frac{ix^{2}\delta_{ab}}{2^{7}\cdot3^{2}}m_{q}\langle g\bar{q}\sigma\cdot Gq\rangle\rlap/x-\frac{x^{4}\delta_{ab}}{2^{10}\cdot3^{3}}\langle\bar{q}q\rangle\langle g^{2}G^{2}\rangle,
\end{eqnarray}
which is then
Fourier-transformed to the momentum space in $D$ dimension.
The
resulting light-quark part is combined with the heavy-quark part
before it is dimensionally regularized at $D=4$.

The spectral densities can be listed as
\begin{eqnarray}
\rho_{i}(s)&=&\rho_{i}^{\mbox{pert}}(s)+\rho_{i}^{\langle\bar{q}q\rangle}(s)+\rho_{i}^{\langle\bar{q}q\rangle^{2}}(s)+
\rho_{i}^{\langle g\bar{q}\sigma\cdot G q\rangle}(s)+\rho_{i}^{\langle
g^{2}G^{2}\rangle}(s)+\rho_{i}^{\langle
g^{3}G^{3}\rangle}(s)+\rho_{i}^{\langle\bar{q}q\rangle^{3}}(s)+
\rho_{i}^{\langle\bar{q}q\rangle\langle
g\bar{q}\sigma\cdot G q\rangle}(s)\nonumber\\&&+
\rho_{i}^{\langle
g\bar{q}\sigma\cdot G q\rangle\langle
g\bar{q}\sigma\cdot G q\rangle}(s)+
\rho_{i}^{\langle\bar{q}q\rangle\langle
g^{2}G^{2}\rangle}(s)+
\rho_{i}^{\langle\bar{q}q\rangle\langle
g^{3}G^{3}\rangle}(s)+
\rho_{i}^{\langle
g^{2}G^{2}\rangle\langle
g\bar{q}\sigma\cdot G q\rangle}(s),~~i=1,2.
\end{eqnarray}
Because many terms of $\rho_{2}(s)$ are
proportional to light quarks' masses and approximate to zero,
we merely present spectral densities
resulted from $\Pi_{1}(q^{2})$.
Concretely, they read
\begin{eqnarray}
\rho_{1}^{\mbox{pert}}(s)&=&\frac{1}{3\cdot5^{2}\cdot2^{16}\pi^{8}}\int_{\Lambda}^{1}d\alpha \frac{(1-\alpha)^{6}}{\alpha^{5}}(\alpha s-m_{Q}^{2})^{4}(\alpha s+4m_{Q}^{2}),\nonumber\\
\rho_{1}^{\langle\bar{q}q\rangle}(s)&=&-\frac{m_{Q}\langle\bar{q}q\rangle}{3\cdot2^{11}\pi^{6}}\int_{\Lambda}^{1}d\alpha\frac{(1-\alpha)^{4}}{\alpha^{3}}(\alpha s-m_{Q}^{2})^{3},\nonumber\\
\rho_{1}^{\langle\bar{q}q\rangle^{2}}(s)&=&\frac{\langle\bar{q}q\rangle^{2}}{3\cdot2^{8}\pi^{4}}\int_{\Lambda}^{1}d\alpha\frac{(1-\alpha)^{3}}{\alpha^{2}}(\alpha s-m_{Q}^{2})(\alpha s+m_{Q}^{2}),\nonumber\\
\rho_{1}^{\langle
g\bar{q}\sigma\cdot G q\rangle}(s)&=&\frac{m_{Q}\langle
g\bar{q}\sigma\cdot G q\rangle}{2^{11}\pi^{6}}\int_{\Lambda}^{1}d\alpha\frac{(1-\alpha)^{3}}{\alpha^{2}}(\alpha s-m_{Q}^{2})^{2},\nonumber\\
\rho_{1}^{\langle
g^{2}G^{2}\rangle}(s)&=&\frac{m_{Q}^{2}\langle
g^{2}G^{2}\rangle}{5\cdot3^{2}\cdot2^{16}\pi^{8}}\int_{\Lambda}^{1}d\alpha\frac{(1-\alpha)^{6}}{\alpha^{5}}(\alpha s-m_{Q}^{2})(\alpha s-2m_{Q}^{2}),\nonumber\\
\rho_{1}^{\langle
g^{3}G^{3}\rangle}(s)&=&\frac{\langle
g^{3}G^{3}\rangle}{5\cdot3^{2}\cdot2^{18}\pi^{8}}\int_{\Lambda}^{1}d\alpha\frac{(1-\alpha)^{6}}{\alpha^{5}}[(\alpha s)^{2}-9\alpha s m_{Q}^{2}+10m_{Q}^{4}],\nonumber\\
\rho_{1}^{\langle\bar{q}q\rangle^{3}}(s)&=&-\frac{m_{Q}\langle\bar{q}q\rangle^{3}}{3\cdot2^{4}\pi^{2}}\int_{\Lambda}^{1}d\alpha(1-\alpha),\nonumber\\
\rho_{1}^{\langle\bar{q}q\rangle\langle
g\bar{q}\sigma\cdot G q\rangle}(s)&=&-\frac{\langle\bar{q}q\rangle\langle
g\bar{q}\sigma\cdot G q\rangle}{2^{8}\pi^{4}}s\int_{\Lambda}^{1}d\alpha(1-\alpha)^{2},\nonumber\\
\rho_{1}^{\langle
g\bar{q}\sigma\cdot G q\rangle\langle
g\bar{q}\sigma\cdot G q\rangle}(s)&=&\frac{\langle
g\bar{q}\sigma\cdot G q\rangle^{2}}{2^{10}\pi^{4}}\int_{\Lambda}^{1}d\alpha(1-\alpha),\nonumber\\
\rho_{1}^{\langle\bar{q}q\rangle\langle
g^{2}G^{2}\rangle}(s)&=&-\frac{m_{Q}\langle\bar{q}q\rangle\langle
g^{2}G^{2}\rangle}{3^{2}\cdot2^{13}\pi^{6}}\int_{\Lambda}^{1}d\alpha\frac{(1-\alpha)^{2}}{\alpha^{3}}[6\alpha^{2}(\alpha s-m_{Q}^{2})+(1-\alpha)^{2}(3\alpha s-4m_{Q}^{2})],\nonumber\\
\rho_{1}^{\langle\bar{q}q\rangle\langle
g^{3}G^{3}\rangle}(s)&=&\frac{m_{Q}\langle\bar{q}q\rangle\langle
g^{3}G^{3}\rangle}{3\cdot2^{14}\pi^{6}}\int_{\Lambda}^{1}d\alpha\frac{(1-\alpha)^{4}}{\alpha^{3}},\nonumber\\
\rho_{1}^{\langle
g^{2}G^{2}\rangle\langle
g\bar{q}\sigma\cdot G q\rangle}(s)&=&\frac{m_{Q}\langle
g^{2}G^{2}\rangle\langle
g\bar{q}\sigma\cdot G q\rangle}{3\cdot2^{13}\pi^{6}}\int_{\Lambda}^{1}d\alpha\frac{(1-\alpha)^{3}}{\alpha^{2}},\nonumber\\
\hat{B}\Pi_{1}^{\mbox{cond}}&=&
\frac{m_{Q}\langle\bar{q}q\rangle^{2}\langle
g\bar{q}\sigma\cdot G q\rangle}{2^{6}\pi^{2}}\int_{0}^{1}d\alpha e^{-m_{Q}^{2}/(\alpha M^{2})}
+\frac{m_{Q}^{2}\langle
g\bar{q}\sigma\cdot G q\rangle^{2}}{2^{10}\pi^{4}}\int_{0}^{1}d\alpha\frac{1-\alpha}{\alpha}e^{-m_{Q}^{2}/(\alpha M^{2})}\nonumber\\&&
-\frac{m_{Q}\langle\bar{q}q\rangle\langle
g^{2}G^{2}\rangle^{2}}{3^{3}\cdot2^{15}\pi^{6}}\int_{0}^{1}d\alpha\frac{(1-\alpha)^{2}}{\alpha}\bigg(\frac{3}{\alpha}-\frac{m_{Q}^{2}}{\alpha^{2}M^{2}}\bigg)e^{-m_{Q}^{2}/(\alpha M^{2})}\nonumber\\&&
-\frac{m_{Q}^{3}\langle\bar{q}q\rangle\langle
g^{3}G^{3}\rangle}{3^{2}\cdot2^{14}\pi^{6}}\int_{0}^{1}d\alpha\frac{(1-\alpha)^{4}}{\alpha^{4}}e^{-m_{Q}^{2}/(\alpha M^{2})}\nonumber\\&&
+\frac{m_{Q}^{2}\langle\bar{q}q\rangle^{2}\langle
g^{2}G^{2}\rangle}{3^{3}\cdot2^{10}\pi^{4}}\int_{0}^{1}d\alpha\frac{(1-\alpha)^{3}}{\alpha^{2}}\bigg(\frac{2}{\alpha}-\frac{m_{Q}^{2}}{\alpha^{2}M^{2}}\bigg)e^{-m_{Q}^{2}/(\alpha M^{2})}\nonumber\\&&
+\frac{\langle\bar{q}q\rangle^{2}\langle
g^{3}G^{3}\rangle}{3^{3}\cdot2^{13}\pi^{4}}\int_{0}^{1}d\alpha\frac{(1-\alpha)^{3}}{\alpha^{2}}\bigg(\frac{11m_{Q}^{2}}{\alpha}-\frac{8m_{Q}^{4}}{\alpha^{2}M^{2}}\bigg)e^{-m_{Q}^{2}/(\alpha M^{2})}\nonumber\\&&
+\frac{\langle\bar{q}q\rangle\langle
g^{2}G^{2}\rangle\langle
g\bar{q}\sigma\cdot G q\rangle}{3^{2}\cdot2^{11}\pi^{4}}\int_{0}^{1}d\alpha\bigg[-1-\frac{(2-4\alpha+3\alpha^{2})m_{Q}^{2}}{\alpha^{3}M^{2}}+\frac{(1-\alpha)^{2}m_{Q}^{4}}{\alpha^{4}(M^{2})^{2}}\bigg]e^{-m_{Q}^{2}/(\alpha M^{2})}\nonumber\\&&
-\frac{m_{Q}^{3}\langle
g^{2}G^{2}\rangle\langle
g\bar{q}\sigma\cdot G q\rangle}{3^{2}\cdot2^{13}\pi^{6}}\int_{0}^{1}d\alpha\frac{(1-\alpha)^{3}}{\alpha^{3}}e^{-m_{Q}^{2}/(\alpha M^{2})}\nonumber\\&&
-\frac{m_{Q}\langle
g\bar{q}\sigma\cdot G q\rangle\langle
g^{3}G^{3}\rangle}{3^{2}\cdot2^{14}\pi^{6}}\int_{0}^{1}d\alpha\frac{(1-\alpha)^{3}}{\alpha^{2}}\bigg(\frac{3}{\alpha}-\frac{m_{Q}^{2}}{\alpha^{2}M^{2}}\bigg)e^{-m_{Q}^{2}/(\alpha M^{2})},\nonumber
\end{eqnarray}
for the $S$-wave $DN$ or $\bar{B}N$ state with $J^{P}=\frac{1}{2}^{-}$, and
\begin{eqnarray}
\rho_{1}^{\mbox{pert}}(s)&=&\frac{1}{3\cdot5^{2}\cdot2^{16}\pi^{8}}\int_{\Lambda}^{1}d\alpha \frac{(1-\alpha)^{6}}{\alpha^{5}}(\alpha s-m_{Q}^{2})^{4}[\alpha s+4m_{Q}^{2}-\alpha(\alpha s-m_{Q}^{2})],\nonumber\\
\rho_{1}^{\langle\bar{q}q\rangle}(s)&=&-\frac{m_{Q}\langle\bar{q}q\rangle}{3\cdot2^{11}\pi^{6}}\int_{\Lambda}^{1}d\alpha\frac{(1-\alpha)^{4}}{\alpha^{3}}(\alpha s-m_{Q}^{2})^{3},\nonumber\\
\rho_{1}^{\langle\bar{q}q\rangle^{2}}(s)&=&\frac{\langle\bar{q}q\rangle^{2}}{3\cdot2^{8}\pi^{4}}\int_{\Lambda}^{1}d\alpha\frac{(1-\alpha)^{3}}{\alpha^{2}}(\alpha s-m_{Q}^{2})[(\alpha s+m_{Q}^{2})-\alpha(\alpha s-m_{Q}^{2})],\nonumber\\
\rho_{1}^{\langle
g\bar{q}\sigma\cdot G q\rangle}(s)&=&\frac{m_{Q}\langle
g\bar{q}\sigma\cdot G q\rangle}{2^{11}\pi^{6}}\int_{\Lambda}^{1}d\alpha\frac{(1-\alpha)^{3}}{\alpha^{2}}(\alpha s-m_{Q}^{2})^{2},\nonumber\\
\rho_{1}^{\langle
g^{2}G^{2}\rangle}(s)&=&\frac{m_{Q}^{2}\langle
g^{2}G^{2}\rangle}{5\cdot3^{2}\cdot2^{17}\pi^{8}}\int_{\Lambda}^{1}d\alpha\frac{(1-\alpha)^{6}}{\alpha^{5}}(\alpha s-m_{Q}^{2})[\alpha(\alpha s-m_{Q}^{2})+2(\alpha s-2m_{Q}^{2})],\nonumber\\
\rho_{1}^{\langle
g^{3}G^{3}\rangle}(s)&=&\frac{\langle
g^{3}G^{3}\rangle}{5\cdot3^{2}\cdot2^{19}\pi^{8}}\int_{\Lambda}^{1}d\alpha\frac{(1-\alpha)^{6}}{\alpha^{5}}[\alpha(\alpha s-m_{Q}^{2})(\alpha s-5m_{Q}^{2})+2((\alpha s)^{2}-9\alpha s m_{Q}^{2}+10m_{Q}^{4})],\nonumber\\
\rho_{1}^{\langle\bar{q}q\rangle^{3}}(s)&=&-\frac{m_{Q}\langle\bar{q}q\rangle^{3}}{3\cdot2^{4}\pi^{2}}\int_{\Lambda}^{1}d\alpha(1-\alpha),\nonumber\\
\rho_{1}^{\langle\bar{q}q\rangle\langle
g\bar{q}\sigma\cdot G q\rangle}(s)&=&\frac{\langle\bar{q}q\rangle\langle
g\bar{q}\sigma\cdot G q\rangle}{2^{8}\pi^{4}}\int_{\Lambda}^{1}d\alpha(1-\alpha)^{2}[(\alpha-1)s-m_{Q}^{2}],\nonumber\\
\rho_{1}^{\langle
g\bar{q}\sigma\cdot G q\rangle\langle
g\bar{q}\sigma\cdot G q\rangle}(s)&=&\frac{\langle
g\bar{q}\sigma\cdot G q\rangle^{2}}{2^{10}\pi^{4}}\int_{\Lambda}^{1}d\alpha(1-\alpha)^{2},\nonumber\\
\rho_{1}^{\langle\bar{q}q\rangle\langle
g^{2}G^{2}\rangle}(s)&=&-\frac{m_{Q}\langle\bar{q}q\rangle\langle
g^{2}G^{2}\rangle}{3^{2}\cdot2^{13}\pi^{6}}\int_{\Lambda}^{1}d\alpha\frac{(1-\alpha)^{2}}{\alpha^{3}}[6\alpha^{2}(\alpha s-m_{Q}^{2})+(1-\alpha)^{2}(3\alpha s-4m_{Q}^{2})],\nonumber\\
\rho_{1}^{\langle\bar{q}q\rangle\langle
g^{3}G^{3}\rangle}(s)&=&\frac{m_{Q}\langle\bar{q}q\rangle\langle
g^{3}G^{3}\rangle}{3\cdot2^{14}\pi^{6}}\int_{\Lambda}^{1}d\alpha\frac{(1-\alpha)^{4}}{\alpha^{3}},\nonumber\\
\rho_{1}^{\langle
g^{2}G^{2}\rangle\langle
g\bar{q}\sigma\cdot G q\rangle}(s)&=&\frac{m_{Q}\langle
g^{2}G^{2}\rangle\langle
g\bar{q}\sigma\cdot G q\rangle}{3\cdot2^{13}\pi^{6}}\int_{\Lambda}^{1}d\alpha\frac{(1-\alpha)^{3}}{\alpha^{2}},\nonumber\\
\hat{B}\Pi_{1}^{\mbox{cond}}&=&
\frac{m_{Q}\langle\bar{q}q\rangle^{2}\langle
g\bar{q}\sigma\cdot G q\rangle}{2^{6}\pi^{2}}\int_{0}^{1}d\alpha e^{-m_{Q}^{2}/(\alpha M^{2})}
+\frac{m_{Q}^{2}\langle
g\bar{q}\sigma\cdot G q\rangle^{2}}{2^{10}\pi^{4}}\int_{0}^{1}d\alpha\frac{1-\alpha}{\alpha}e^{-m_{Q}^{2}/(\alpha M^{2})}\nonumber\\&&
-\frac{m_{Q}\langle\bar{q}q\rangle\langle
g^{2}G^{2}\rangle^{2}}{3^{3}\cdot2^{15}\pi^{6}}\int_{0}^{1}d\alpha\frac{(1-\alpha)^{2}}{\alpha}\bigg(\frac{3}{\alpha}-\frac{m_{Q}^{2}}{\alpha^{2}M^{2}}\bigg)e^{-m_{Q}^{2}/(\alpha M^{2})}\nonumber\\&&
-\frac{m_{Q}^{3}\langle\bar{q}q\rangle\langle
g^{3}G^{3}\rangle}{3^{2}\cdot2^{14}\pi^{6}}\int_{0}^{1}d\alpha\frac{(1-\alpha)^{4}}{\alpha^{4}}e^{-m_{Q}^{2}/(\alpha M^{2})}\nonumber\\&&
-\frac{m_{Q}^{2}\langle\bar{q}q\rangle^{2}\langle
g^{2}G^{2}\rangle}{3^{3}\cdot2^{10}\pi^{4}}\int_{0}^{1}d\alpha\frac{(1-\alpha)^{3}}{\alpha^{3}}\bigg[-(\alpha+2)+\frac{m_{Q}^{2}}{\alpha M^{2}}\bigg]e^{-m_{Q}^{2}/(\alpha M^{2})}\nonumber\\&&
-\frac{\langle\bar{q}q\rangle^{2}\langle
g^{3}G^{3}\rangle}{3^{3}\cdot2^{13}\pi^{4}}\int_{0}^{1}d\alpha\frac{(1-\alpha)^{3}}{\alpha^{3}}\bigg[-(11m_{Q}^{2}+2\alpha)+\frac{4m_{Q}^{2}(2m_{Q}^{2}+\alpha)}{\alpha M^{2}}\bigg]e^{-m_{Q}^{2}/(\alpha M^{2})}\nonumber\\&&
-\frac{\langle\bar{q}q\rangle\langle
g^{2}G^{2}\rangle\langle
g\bar{q}\sigma\cdot G q\rangle}{3^{2}\cdot2^{11}\pi^{4}}\int_{0}^{1}d\alpha\bigg[(1-\alpha)+\frac{(\alpha^{2}+(\alpha+2)(1-\alpha)^{2})m_{Q}^{2}}{\alpha^{3}M^{2}}-\frac{(1-\alpha)^{2}m_{Q}^{4}}{\alpha^{4}(M^{2})^{2}}\bigg]e^{-m_{Q}^{2}/(\alpha M^{2})}\nonumber\\&&
-\frac{m_{Q}^{3}\langle
g^{2}G^{2}\rangle\langle
g\bar{q}\sigma\cdot G q\rangle}{3^{2}\cdot2^{13}\pi^{6}}\int_{0}^{1}d\alpha\frac{(1-\alpha)^{3}}{\alpha^{3}}e^{-m_{Q}^{2}/(\alpha M^{2})}\nonumber\\&&
-\frac{m_{Q}\langle
g\bar{q}\sigma\cdot G q\rangle\langle
g^{3}G^{3}\rangle}{3^{2}\cdot2^{14}\pi^{6}}\int_{0}^{1}d\alpha\frac{(1-\alpha)^{3}}{\alpha^{2}}\bigg(\frac{3}{\alpha}-\frac{m_{Q}^{2}}{\alpha^{2}M^{2}}\bigg)e^{-m_{Q}^{2}/(\alpha M^{2})},\nonumber
\end{eqnarray}
for the $S$-wave $D^{*}N$ or $\bar{B}^{*}N$ state with $J^{P}=\frac{3}{2}^{-}$.
 The lower limit of  integration is
given by $\Lambda=m_{Q}^{2}/s$.
In the deriving of above results,
we have applied the factorization hypothesis of the four quark condensate
$\langle q\bar{q}q\bar{q}\rangle=\kappa\langle\bar{q}q\rangle\langle\bar{q}q\rangle$
and have set $\kappa=1$ following the usual treatment.
Numerically, the factor $\kappa$ may have some other value such as $2$
or $3$.

\section{Numerical analysis and discussions}\label{sec3}
We perform numerical analysis of the sum rule
(\ref{sumrule1}) to extract mass values of studied states.
One could take input parameters as
$m_{c}=1.23\pm0.05~\mbox{GeV}$, $m_{b}=4.24\pm0.06~\mbox{GeV}$,
$\langle\bar{q}q\rangle=-(0.23\pm0.03)^{3}~\mbox{GeV}^{3}$, $\langle
g\bar{q}\sigma\cdot G q\rangle=m_{0}^{2}~\langle\bar{q}q\rangle$,
$m_{0}^{2}=0.8\pm0.1~\mbox{GeV}^{2}$, $\langle
g^{2}G^{2}\rangle=0.88~\mbox{GeV}^{4}$, and $\langle
g^{3}G^{3}\rangle=0.045~\mbox{GeV}^{6}$
\cite{overview2}.
To choose proper work windows for the threshold $s_0$ and the Borel parameter $M^2$,
one could consider two rules in the standard QCD approach:
on one hand, the
perturbative contribution should be larger than condensate
contributions to have a good
convergence in the
OPE side; on the other
hand, the pole contribution should be larger than the continuum
state contributions
in the phenomenological
side. Besides the above two restrictions, the threshold parameter
$\sqrt{s_{0}}$ should not be taken arbitrarily.
It is known that the first excitation of
studied state defines the size of $\sqrt{s_0}$, and $\sqrt{s_0}$
should be higher than the extracted value $M_{H}$ of studied state around
$0.5~\mbox{GeV}$ for many hadrons.
Taking the case of $DN$ state as an example, if $\sqrt{s_0}$
were taken as $3.2\sim3.4~\mbox{GeV}$,
one could obtain the mass $M_{H}=3.52\pm0.36~\mbox{GeV}$.
However, the value of $\sqrt{s_0}-M_{H}$
is too small or even minus,
which means that the values of continuum
threshold $\sqrt{s_0}$ are taken a bit small and should be increased correspondingly.

However, it may have some difficulty to find a conventional work window critically satisfying all
the above rules in this work, which has been discussed in detail for some other cases, e.g. Refs. \cite{HXChen,ZGWang,Matheus,Zs}.
The main reason is that some condensate contributions are very large,
making the standard OPE convergence (i.e. perturbative contribution at least
larger than each condensate contribution) to happen only at very large values of $M^2$.
For the case of $S$-wave $DN$ state with $J^{P}=\frac{1}{2}^{-}$ as an example,
the comparison
between pole and continuum contributions from the
sum rule (\ref{sumrule1}) for $\sqrt{s_0}=4.0~\mbox{GeV}$
 is shown in the left panel of Fig. 1, and its OPE
convergence is shown by comparing the perturbative with
other condensate contributions in the right panel.
One can see that there are four main
condensates (i.e., $\langle\bar{q}q\rangle$, $\langle g\bar{q}\sigma\cdot G q\rangle$,
$\langle\bar{q}q\rangle^{2}$, and $\langle\bar{q}q\rangle\langle
g\bar{q}\sigma\cdot G q\rangle$), and they could cancel each other out to some extent
as they have different signs. Besides, most of the other
condensates calculated are very small and almost negligible.
Thus, one could try releasing the rigid OPE convergence
criterion (i.e., that the perturbative contribution should
be larger than each condensate contribution) and restrict
the ratio of the perturbative to the ``total OPE contribution"
(the sum of the perturbative and other condensates
calculated) to be at least larger than one half or more. What is also very important
that we have found that condensates higher than
dimension $12$ are quite small, and they
could not radically influence the character of OPE convergence
here. All these factors
bring that the OPE convergence is still under control at relatively low values of $M^{2}$.
The dependence on Borel parameter $M^{2}$ for
masses of $S$-wave $DN$ and $\bar{B}N$ states with $J^{P}=\frac{1}{2}^{-}$ are shown in FIG. 2, for which continuum
thresholds are taken as $\sqrt{s_0}=3.9\sim4.1~\mbox{GeV}$ and $\sqrt{s_0}=7.4\sim7.6~\mbox{GeV}$, respectively.
From the Borel curves, one can visually see that there indeed exist stable plateaus.
Thus,
we choose some transition range $M^{2}=2.0\sim3.0~\mbox{GeV}^{2}$ as a compromise Borel window
and arrive at
$3.75\pm0.14~\mbox{GeV}$ for $DN$ state.
Considering the uncertainty rooting in the variation of quark masses and
condensates, we gain
$3.75\pm0.14\pm0.08~\mbox{GeV}$ (the
first error reflects the uncertainty due to variation of $\sqrt{s_{0}}$
and $M^{2}$, and the second error resulted from the variation of
QCD parameters)
for the $S$-wave $DN$ state with $J^{P}=\frac{1}{2}^{-}$.
To investigate the effect of different factorization, we take $\kappa=2$
and arrive at $3.56\pm0.10\pm0.07~\mbox{GeV}$ from the similar analysis
process. Similarly, the result is $3.45\pm0.07\pm0.07~\mbox{GeV}$ while $\kappa=3$.
Finally, we average three results
for $\kappa=1\sim3$ and arrive at the mass value $3.64\pm0.33~\mbox{GeV}$ concisely for the
$S$-wave $DN$ state with $J^{P}=\frac{1}{2}^{-}$,
which is somewhat higher than the experimental value of $\Sigma_{c}(2800)$ even considering the uncertainty of result.
For the $S$-wave $\bar{B}N$ state with $J^{P}=\frac{1}{2}^{-}$,
we choose some transition range $M^{2}=4.0\sim5.0~\mbox{GeV}^{2}$ as a compromise Borel window
and arrive at
$7.06\pm0.13~\mbox{GeV}$.
Considering the uncertainty rooting in the variation of quark masses and
condensates, we gain
$7.06\pm0.13\pm0.12~\mbox{GeV}$
for the $S$-wave $\bar{B}N$ state with $J^{P}=\frac{1}{2}^{-}$.
The respective results are
$6.91\pm0.10\pm0.10~\mbox{GeV}$ and $6.82\pm0.09\pm0.09~\mbox{GeV}$
for $\kappa=2$ and $3$.
Averaging three results
for $\kappa=1\sim3$, one can arrive at the final mass value $6.97\pm0.34~\mbox{GeV}$ in a nutshell for the
$S$-wave $\bar{B}N$ state with $J^{P}=\frac{1}{2}^{-}$.

\begin{figure}
\centerline{\epsfysize=5.8truecm
\epsfbox{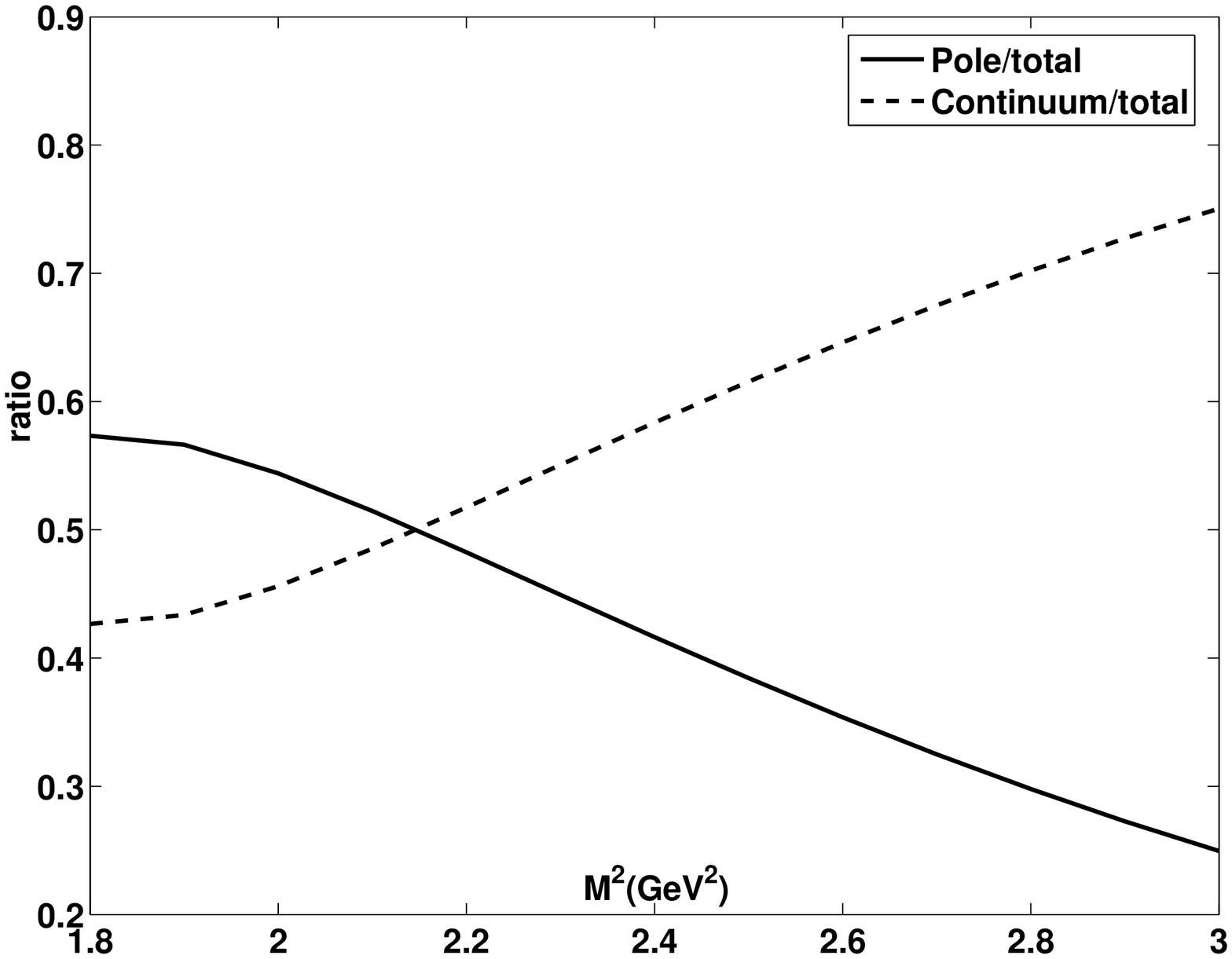}\epsfysize=5.8truecm\epsfbox{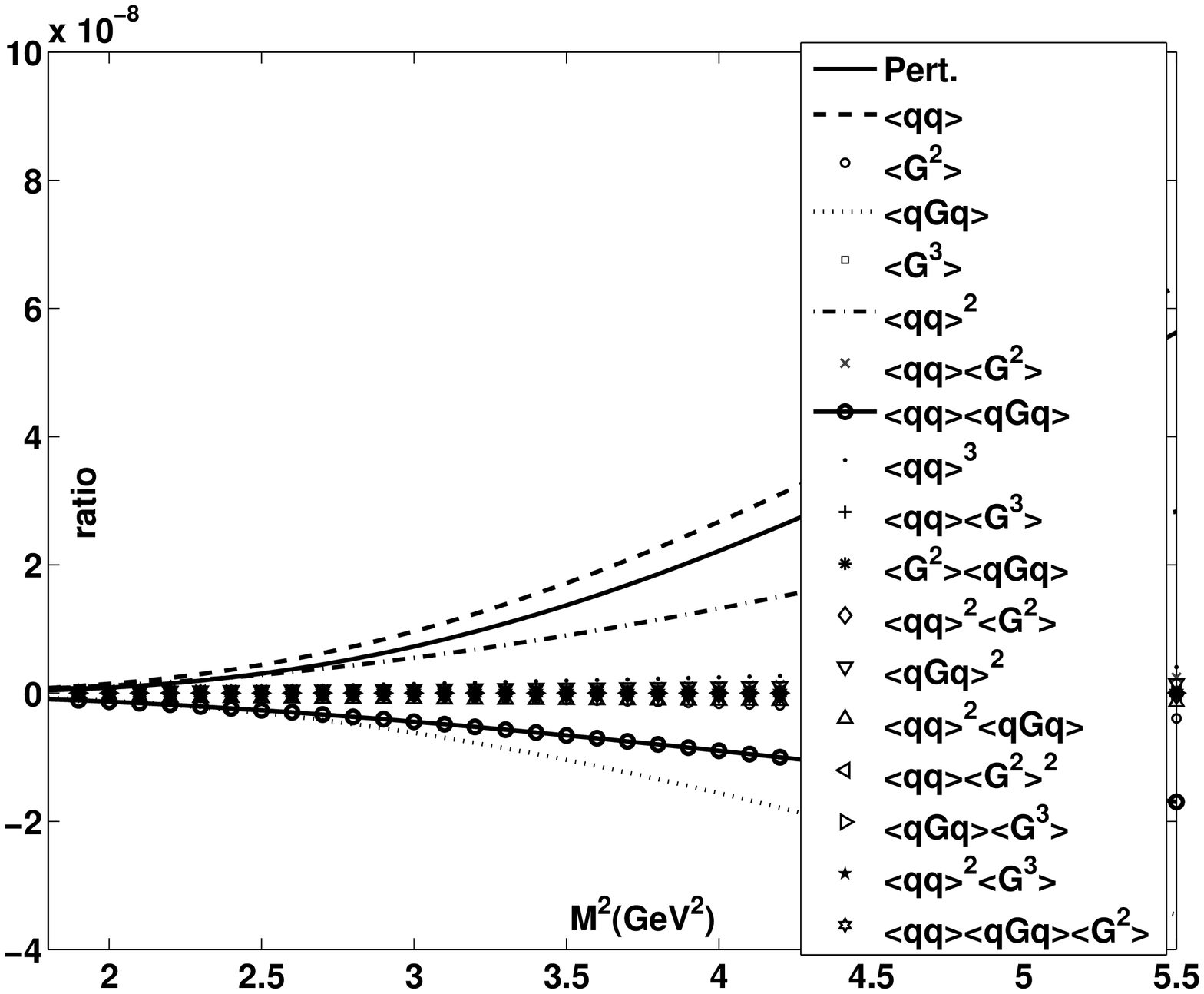}}\caption{In the left panel, the solid line shows the relative pole contribution (the pole contribution divided by the total, pole plus continuum contribution), and the dashed line shows the relative continuum contribution from the sum rule [Eq. (\ref{qsr1})] for $\sqrt{s_0}=4.0~\mbox{GeV}$ for the $S$-wave $DN$ state with $J^{P}=\frac{1}{2}^{-}$. In the right panel, the OPE convergence is shown by comparing the perturbative with other condensate contributions from the sum rule [Eq. (\ref{qsr1})] for $\sqrt{s_0}=4.0~\mbox{GeV}$ for the $S$-wave $DN$ state with $J^{P}=\frac{1}{2}^{-}$.}
\end{figure}

\begin{figure}
\centerline{\epsfysize=5.8truecm
\epsfbox{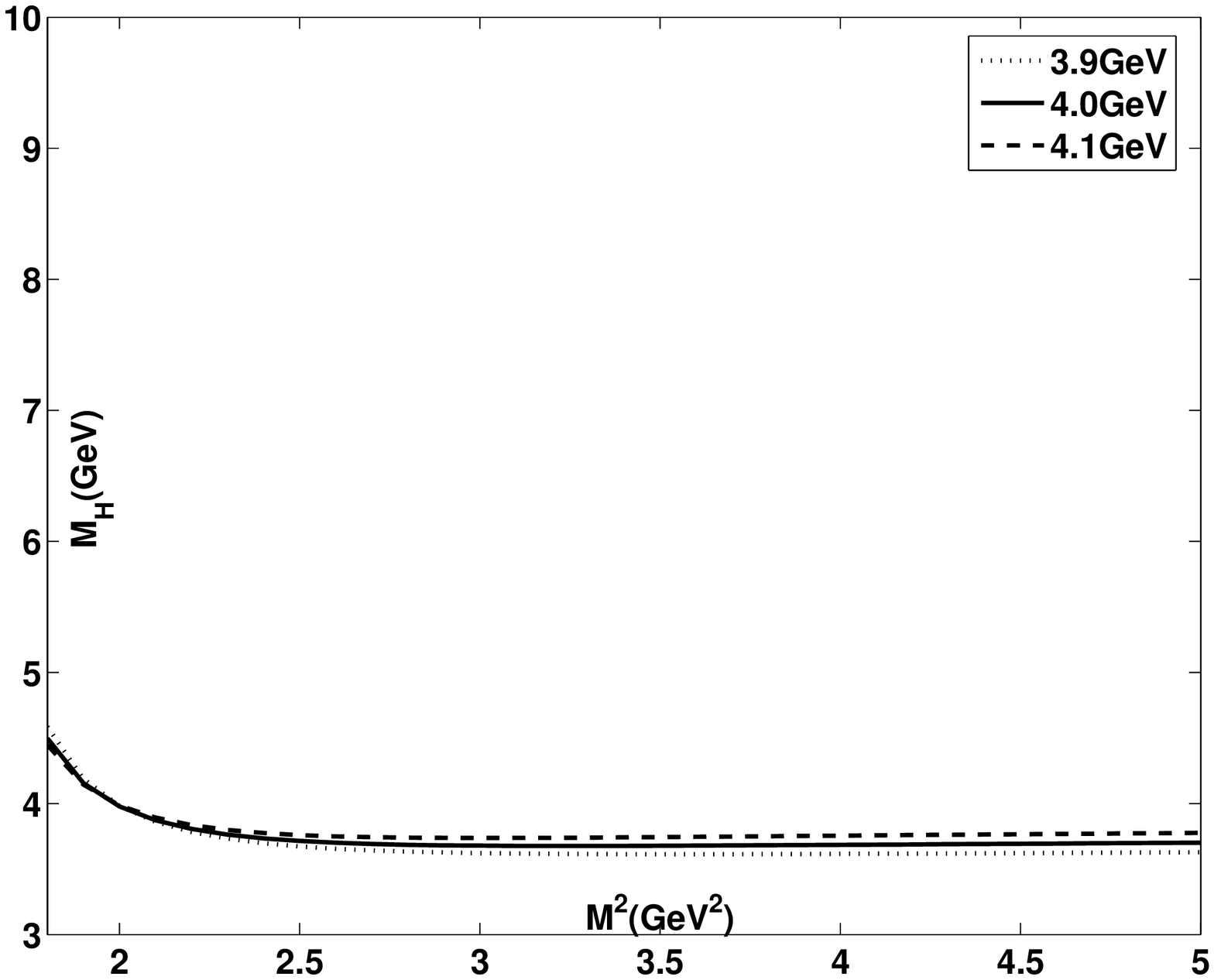}\epsfysize=5.8truecm\epsfbox{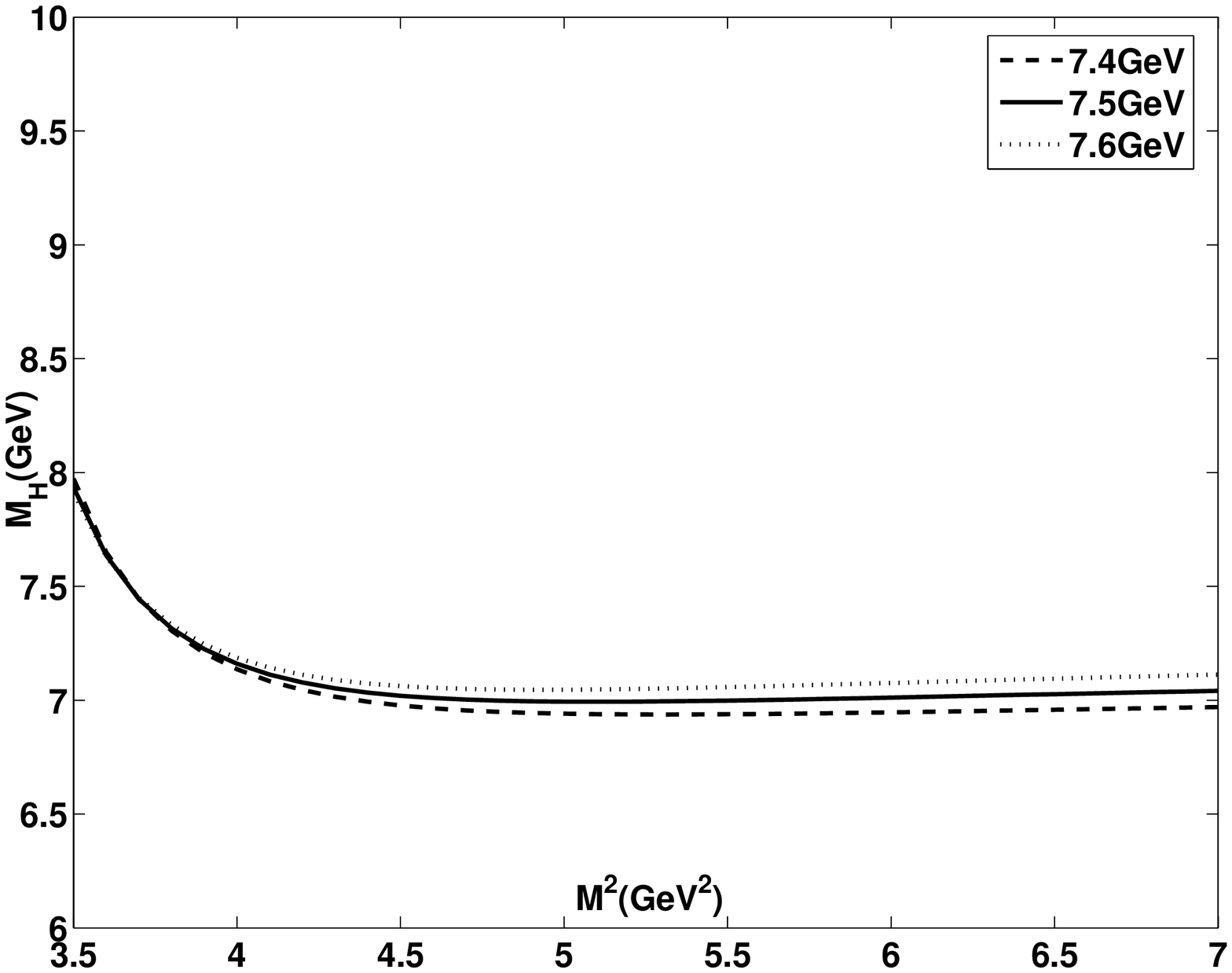}}\caption{Masses of $S$-wave $DN$ and $\bar{B}N$ states with $J^{P}=\frac{1}{2}^{-}$ as a function of $M^{2}$ from the sum rule [Eq. (\ref{sumrule1})] are shown. The continuum
thresholds are taken as $\sqrt{s_0}=3.9\sim4.1~\mbox{GeV}$ and $\sqrt{s_0}=7.4\sim7.6~\mbox{GeV}$, respectively.}
\end{figure}

For another example, the comparison
between pole and continuum contributions from the
sum rule (\ref{sumrule1}) for $\sqrt{s_0}=4.1~\mbox{GeV}$ for the $S$-wave $D^{*}N$ state with $J^{P}=\frac{3}{2}^{-}$
 is shown in the left panel of Fig. 3, and its OPE
convergence  is shown by comparing the perturbative with
other condensate contributions  in the right panel.
Masses of $S$-wave $D^{*}N$ and $\bar{B}^{*}N$ states with $J^{P}=\frac{3}{2}^{-}$ as a function of $M^{2}$ from sum rule
(\ref{sumrule1}) are
shown in FIG. 4.
Graphically, one can see that there have stable plateaus for the Borel curves.
Similarly,
we choose some transition range $M^{2}=2.0\sim3.0~\mbox{GeV}^{2}$ as a compromise Borel window for $D^{*}N$ state,
and arrive at
$3.83\pm0.16~\mbox{GeV}$.
Considering the uncertainty rooting in the variation of quark masses and
condensates, we obtain
$3.83\pm0.16\pm0.09~\mbox{GeV}$
for the $S$-wave $D^{*}N$ state with $J^{P}=\frac{3}{2}^{-}$.
Taking $\kappa=2$ and $3$, the results are
$3.62\pm0.09\pm0.07~\mbox{GeV}$ and $3.52\pm0.07\pm0.07~\mbox{GeV}$, respectively.
Averaging three results for $\kappa=1\sim3$, the final result is
$3.73\pm0.35~\mbox{GeV}$ in a concise form for the
$S$-wave $D^{*}N$ state with $J^{P}=\frac{3}{2}^{-}$,
which is bigger than the experimental data of $\Lambda_{c}(2940)^{+}$
even taking into account the uncertainty.
For the $S$-wave $\bar{B}^{*}N$ state with $J^{P}=\frac{3}{2}^{-}$,
we choose a compromise Borel window $M^{2}=4.0\sim5.0~\mbox{GeV}^{2}$
and take $\sqrt{s_0}=7.4\sim7.6~\mbox{GeV}$.
In the work windows, we obtain
$7.07\pm0.12~\mbox{GeV}$ for $\bar{B}^{*}N$ state. Varying input values of quark masses and
condensates, we attain
$7.07\pm0.12\pm0.12~\mbox{GeV}$
for the $S$-wave $\bar{B}^{*}N$ state with $J^{P}=\frac{3}{2}^{-}$.
The results are
$6.92\pm0.11\pm0.10~\mbox{GeV}$ and $6.83\pm0.10\pm0.09~\mbox{GeV}$ for $\kappa=2$ and $\kappa=3$,
respectively.
Making the average of three results for $\kappa=1\sim3$,
one could gain
$6.98\pm0.34~\mbox{GeV}$ concisely for the
$S$-wave $\bar{B}^{*}N$ state with $J^{P}=\frac{3}{2}^{-}$.

\begin{figure}
\centerline{\epsfysize=5.8truecm
\epsfbox{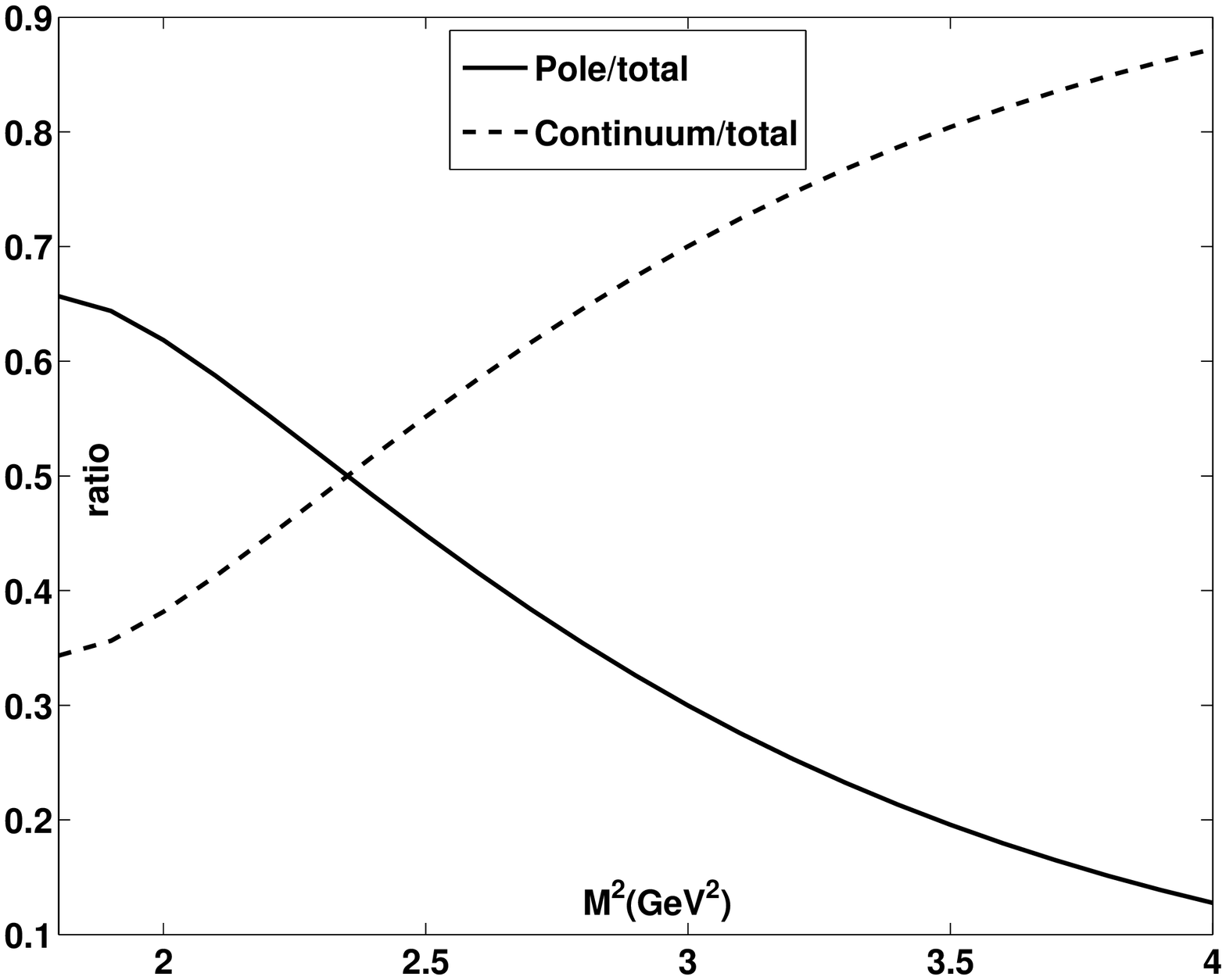}\epsfysize=5.8truecm\epsfbox{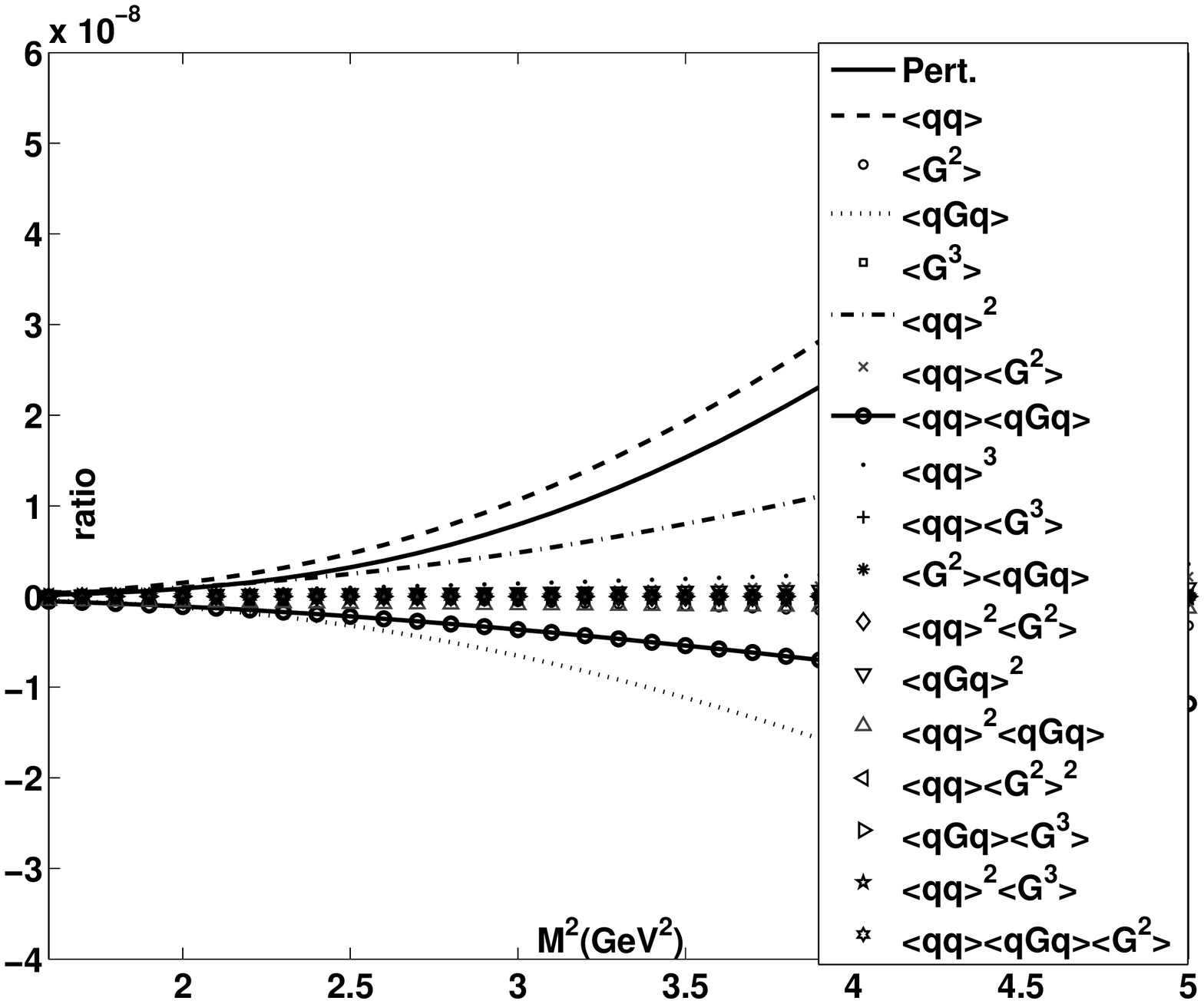}}\caption{In the left panel, the solid line shows the relative pole contribution (the pole contribution divided by the total, pole plus continuum contribution), and the dashed line shows the relative continuum contribution from the sum rule [Eq. (\ref{qsr1})] for $\sqrt{s_0}=4.1~\mbox{GeV}$ for the $S$-wave $D^{*}N$ state with $J^{P}=\frac{3}{2}^{-}$. In the right panel, the OPE convergence is shown by comparing the perturbative with other condensate contributions from the sum rule [Eq. (\ref{qsr1})] for $\sqrt{s_0}=4.1~\mbox{GeV}$ for the $S$-wave $D^{*}N$ state with $J^{P}=\frac{3}{2}^{-}$.}
\end{figure}

\begin{figure}
\centerline{\epsfysize=5.8truecm
\epsfbox{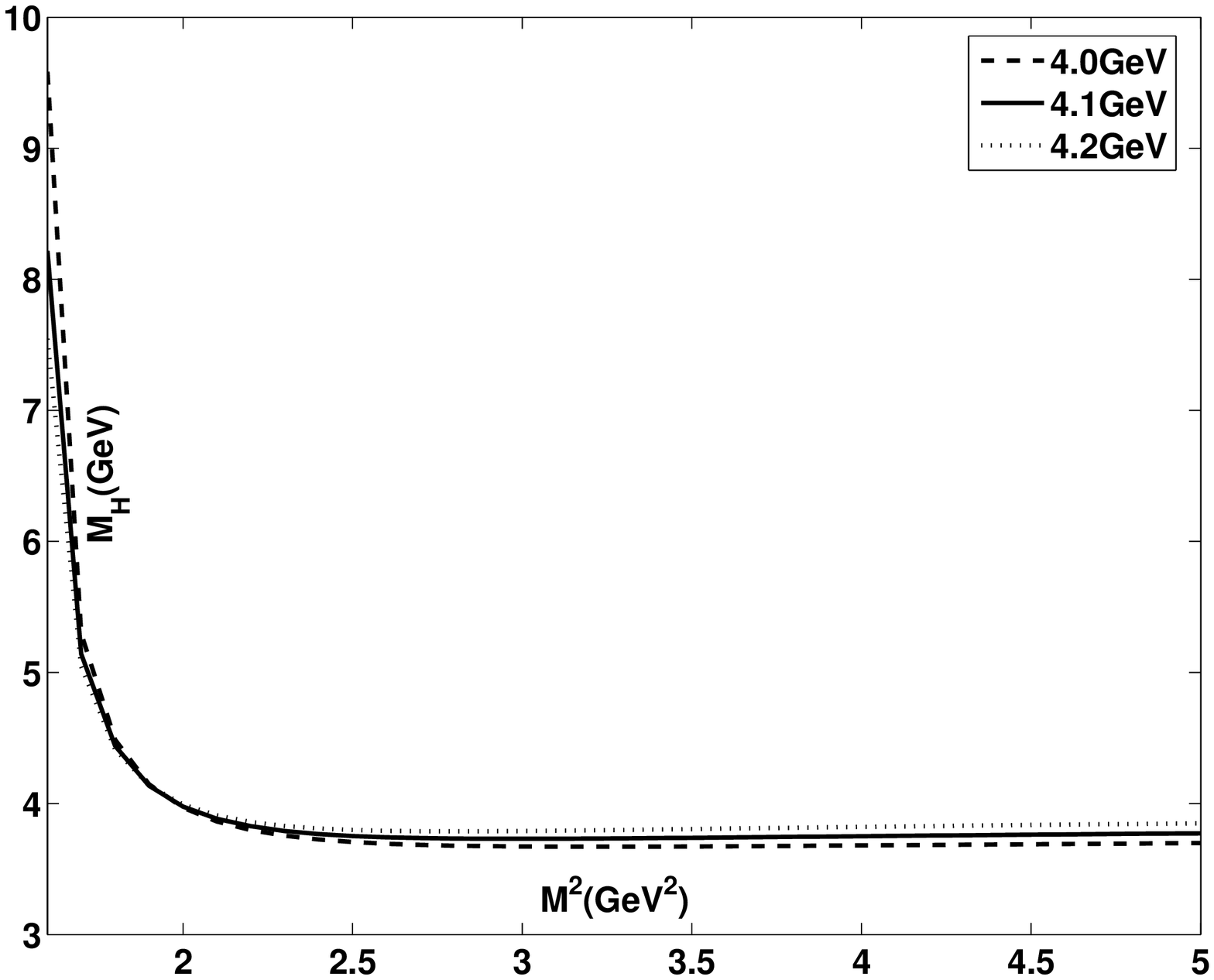}\epsfysize=5.8truecm
\epsfbox{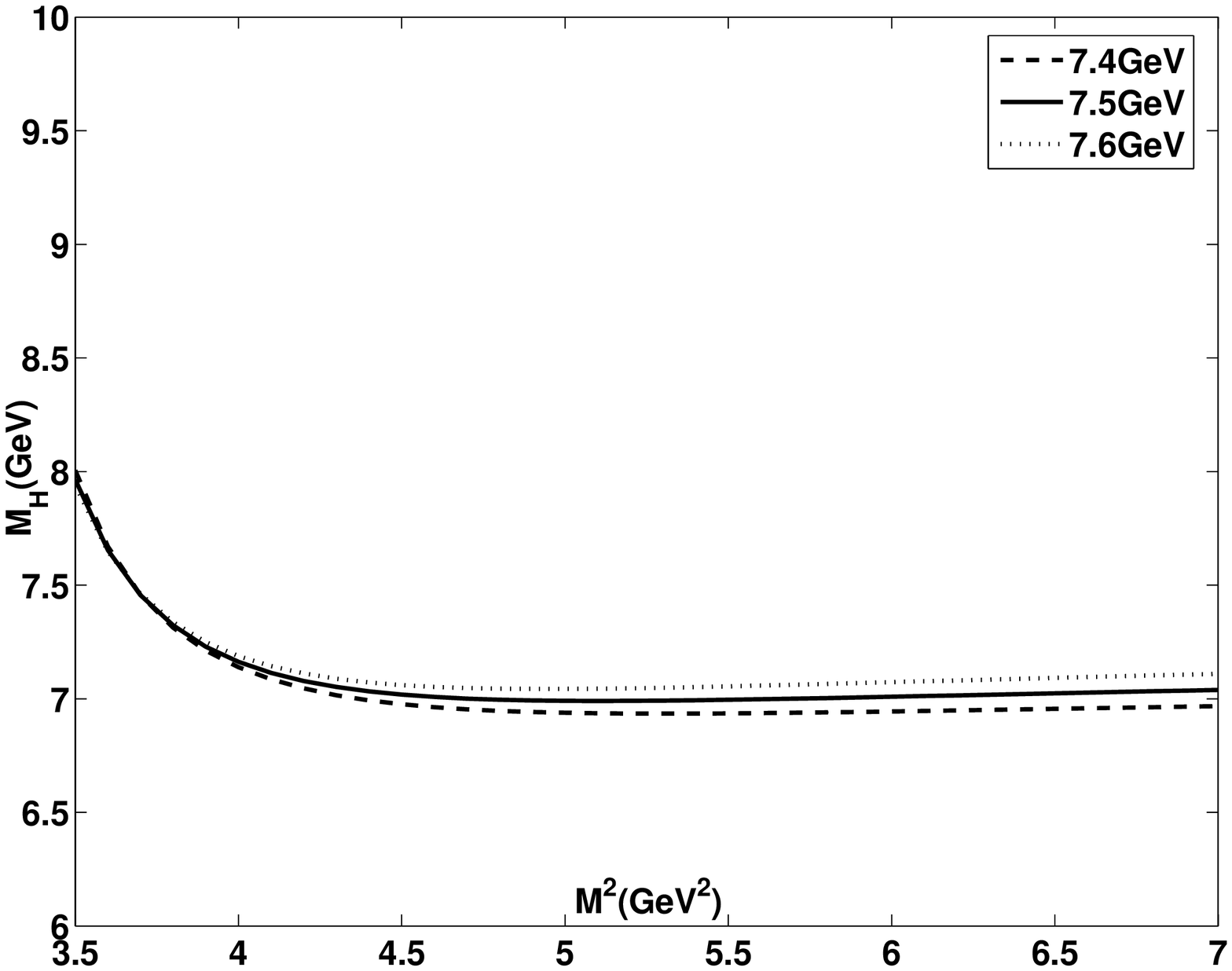}}\caption{Masses of $S$-wave $D^{*}N$ and $\bar{B}^{*}N$ states with $J^{P}=\frac{3}{2}^{-}$ as a function of $M^{2}$ from the sum rule [Eq. (\ref{sumrule1})] are shown. The continuum
thresholds are taken as $\sqrt{s_0}=4.0\sim4.2~\mbox{GeV}$ and $\sqrt{s_0}=7.4\sim7.6~\mbox{GeV}$, respectively.}
\end{figure}

\section{Conclusions}\label{sec4}
In some theoretical approaches, the hadronic
resonances $\Sigma_{c}(2800)$ and $\Lambda_{c}(2940)^{+}$ have been suggested to be $S$-wave
$DN$ and $D^{*}N$ molecular states, respectively.
From QCD sum rules, we investigate that whether $\Sigma_{c}(2800)$ and $\Lambda_{c}(2940)^{+}$
could be the $S$-wave $DN$ state with $J^{P}=\frac{1}{2}^{-}$ and the $S$-wave $D^{*}N$ state with $J^{P}=\frac{3}{2}^{-}$,
respectively.
In the
OPE calculation, contributions of operators up to dimension $12$ are included and one could find
that its convergence is still under control.
The final result for the $S$-wave $DN$ state of $J^{P}=\frac{1}{2}^{-}$ is $3.64\pm0.33~\mbox{GeV}$,
which is somewhat bigger than the experimental value of $\Sigma_{c}(2800)$ even considering the uncertainty of result.
The numerical result for the $S$-wave $D^{*}N$ state of $J^{P}=\frac{3}{2}^{-}$
is $3.73\pm0.35~\mbox{GeV}$, which is a bit higher than the experimental data of $\Lambda_{c}(2940)^{+}$
even taking into account the uncertainty.
Considering that corresponding molecular currents are constructed from local operators,
one can not arbitrarily exclude the possibility that $\Sigma_c(2800)$ and $\Lambda_{c}(2940)^{+}$ are molecular states
just from these disagreements.
However, one can infer that $\Sigma_{c}(2800)$ and $\Lambda_{c}(2940)^{+}$ could not be compact states from these results.
This may suggest a limitation of the QCD sum rule
using the local current to determine whether some state is a molecular state or not.
By the way, we also study the corresponding
bottom counterparts and predict their masses
to be $6.97\pm0.34~\mbox{GeV}$ for the $S$-wave $\bar{B}N$ state with $J^{P}=\frac{1}{2}^{-}$ and
$6.98\pm0.34~\mbox{GeV}$ for the $S$-wave $\bar{B}^{*}N$ state with $J^{P}=\frac{3}{2}^{-}$,
which could be searched in future experiments.
\begin{acknowledgments}
The author would like to
thank Prof. C.~Garc\'{\i}a-Recio and Prof. L.~L.~Salcedo for communications and
discussions.
This work was supported by the National Natural Science
Foundation of China under Contracts No. 11105223, No. 11275268, and the
project in NUDT for excellent youth talents.
\end{acknowledgments}

\end{document}